\documentclass[conference]{IEEEtran}
\IEEEoverridecommandlockouts

\usepackage{cite}
\usepackage{amsmath,amssymb,amsfonts}
\usepackage{algorithmic}
\usepackage{graphicx}
\usepackage{textcomp}
\usepackage{xcolor}

\usepackage{tikz}
\usepackage{wrapfig}
\usepackage{array}
\usepackage{todonotes}
\usepackage{booktabs}
\usepackage{colortbl}
\usepackage{multirow}
\usepackage{enumitem}
\usepackage[ruled,vlined]{algorithm2e}
\usepackage{float}
 \usepackage{url}

\def\BibTeX{{\rm B\kern-.05em{\sc i\kern-.025em b}\kern-.08em
    T\kern-.1667em\lower.7ex\hbox{E}\kern-.125emX}}

\begin{document}

\title{Computational Onboard Data Management for Heterogeneous Autonomous Vehicle Streams}

\author{\IEEEauthorblockN{Yuxin Wang\IEEEauthorrefmark{1}, Yuankai He\IEEEauthorrefmark{1}, Damien Kah\IEEEauthorrefmark{2}, and Weisong Shi\IEEEauthorrefmark{1}}
\IEEEauthorblockA{\IEEEauthorrefmark{1}University of Delaware, Newark, DE, USA}
\IEEEauthorblockA{\IEEEauthorrefmark{2}Western Digital Corporation, San Jose, CA, USA}
}

\maketitle

\begin{abstract}
Autonomous vehicles continuously generate large volumes of heterogeneous sensor and state data streams. While this data supports real-time autonomy, it also enables emerging vehicle computing applications that require retained historical context. Existing in-vehicle logging stacks are largely replay- or compliance-oriented: rich sensor streams are often discarded after online processing or recorded only in limited windows for offline replay. This creates a new onboard vehicle-data operating point: transforming transient multimodal streams into compact, long-horizon, queryable history under embedded resource constraints.

We present Autonomous Vehicle Storage (AVS), a computational and hierarchical in-vehicle data management system for this operating point. AVS introduces a vehicle-specialized architecture including (i) utility-aware modality preprocessing before persistence, (ii) a chunk-indexed append-only logger for sustained ingest, time-range lookup, and prefix recovery, and (iii) an SSD–HDD hot/cold hierarchy that supports direct cold queries. Evaluated on a real L4 platform with three days of traces and concurrent LiDAR, camera, and GPS logging, AVS reduces storage by 8.0-8.7× relative to raw rosbag2 from its modality and task aware preprocessing, delivers hot-tier (SSD) query Time To First Byte (TTFB) of 0.9-9.1 ms and cold-tier (HDD) query TTFB of 25.6-44.0 ms, recovers from power loss in 9.98 s with a bounded 0.24 s loss window, and sustains background archival to HDD at about 107-108 MB/s. It shows the feasibility of AVS on real in-vehicle platforms, and it also exposes key tradeoffs and open challenges for vehicle data engineering.
\end{abstract}


\section{Introduction}
Modern autonomous vehicles (AVs) integrate heterogeneous sensors and substantial onboard compute to sustain real-time perception and planning control loops\cite{liu2020computing}. Beyond these real-time driving functions, vehicles are increasingly viewed as mobile edge computing platforms that also support third-party applications and services, a paradigm often referred to as vehicle computing \cite{lu2023vehicle}. This shift changes the role of onboard data: multimodal streams that were originally consumed transiently by the driving stack must also be retained as queryable history for later applications. Representative examples include battery state of health estimation from fleet traces \cite{theissler2021predictive}, high definition map maintenance through crowdsourced updates \cite{kim2021hdmapupdate},  trajectory data management and analytics~\cite{Ding2018UlTraMan, Li2020TrajMesa} and long horizon telematics analysis for predictive maintenance \cite{theissler2021predictive}. These use cases require retained, queryable data collected during regular vehicle operation.

However, current production vehicle storage practices are largely oriented toward replay, diagnostics, or regulatory logging rather than general-purpose historical access. Commercial recorders typically retain limited and sparse event data for compliance or service workflows. For example, event data recorders capture limited pre- and post-crash windows, recently extended to 20 seconds at 10 Hz \cite{nhtsa2024edr}. These mechanisms are valuable, but they do not expose a general-purpose query interface over heterogeneous sensor histories for the emerging onboard applications.

At the same time, full fidelity AV sensor workloads are large and diverse. Continuous logging of cameras, LiDAR, radar, and vehicle Controller Area Network (CAN) bus can reach terabytes per day~\cite{wang2024quantitative}. Bulk uplink of raw data is therefore constrained by cellular bandwidth and cloud transfer cost, motivating selective retention and processing directly on the vehicle. Industry analyses further note the shift toward in-vehicle inference to meet latency, availability, privacy, and data traffic constraints~\cite{mckinsey_edge_ai_automotive}. Taken together, these trends suggest an objective that differs from replay-oriented logging: retaining historical data on the vehicle while keeping it queryable under constrained resources.

In this paper, we explore this new onboard vehicle-data operating point through the Autonomous Vehicle Storage (AVS), an in-vehicle data management system for vehicle computing workloads. AVS runs on a separate computing unit from the AV main computer to avoid interfering with driving and safety-critical tasks, and it complements rather than replaces existing replay-oriented recorders by targeting sustained stream ingest, queryable retention, and onboard archival. We evaluate AVS in two ways: first, we use task-level datasets with ground truth to choose reduction and compression operators that preserve downstream utility; second, we evaluate the complete stream-to-queryable-history path on a real L4 vehicle to test whether the design can sustain realistic multimodal ingest, recovery, archival, and query workloads, and compare it against rosbag2 as a widely used replay-oriented recording pipeline to quantify end-to-end pipeline behavior under identical input streams, while separately attributing footprint reduction to preprocessing and ingest/query behavior to the storage layout.

The contribution of this paper is as follows, along with its section roadmap:

\begin{itemize}[nosep,leftmargin=*,labelsep=0.5em]
  \item We identify a new onboard vehicle-data operating point between transient stream processing and replay-oriented logging. We formulate the resulting data management requirements for long-horizon AV history retention. (Section~\ref{paper:bg}).

  \item We design \emph{AVS}, a computational and hierarchical in-vehicle data management architecture for transforming heterogeneous AV streams into compact, long-horizon, queryable onboard histories. And we introduce a vehicle-specialized append-only logger as the core storage layout of AVS. (Section~\ref{paper:design} and~\ref{paper:implementation}).

  \item We implement and evaluate an AVS prototype on an edge-deployable Raspberry Pi 5 platform, demonstrating end-to-end behavior across ingest, power-loss recovery, archival, hot/cold query, and selective cloud offload using three days of traces from a real Level-4 driving platform.  (Section~\ref{paper:evaluation}).
\end{itemize}

We summarized the findings for AV data management in the vehicle computing era and discussed future directions and open challenges in Section~\ref{paper:discussion}.
\section{Motivation and Gap Analysis}
\label{paper:bg}
\paragraph{Vehicle Computing Storage Needs}
\label{bg:av-needs}

\begin{figure}[!ht] 
\centering 
\includegraphics[width=\linewidth]{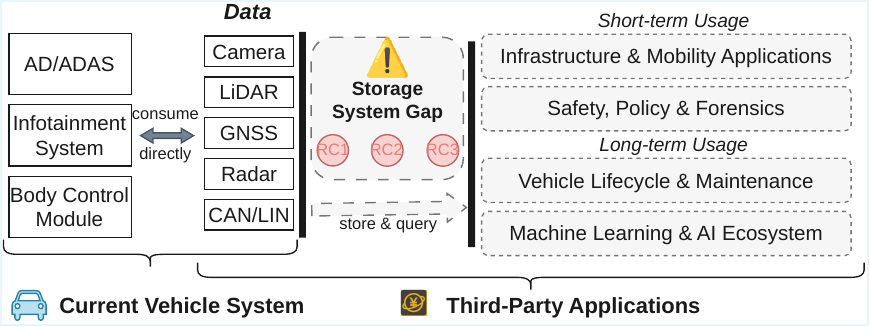} 
\caption{Autonomous vehicle storage gap in the vehicle-computing paradigm.} 
\label{fig:RQ} 
\end{figure}

Figure~\ref{fig:RQ} highlights a gap between how in-vehicle software consumes sensor streams and what emerging vehicle computing applications expect from onboard storage. Today’s AV stacks are largely organized around real-time consumers, including AV and Advanced Driver Assistance Systems(ADAS) functions, infotainment, and control. These components ingest high-rate streams from cameras, LiDAR, Global Navigation Satellite System (GNSS), radar, and CAN bus to support immediate perception and actuation. In this execution model, sensor data are primarily input to online computation, and persistent storage is often limited to constrained logging designed for diagnostics or for specific regulated event records. For example, NHTSA’s event data recorder requirements specify what must be recorded and for how long around a crash event, reinforcing that some onboard records are purpose-scoped and format-constrained rather than designed for general historical data query \cite{nhtsa2024edr}.

Vehicle computing shifts this expectation from narrow logging toward services that operate over historical onboard data. Prior work frames connected vehicles as mobile computing platforms that support third-party applications and services beyond the driving loop~\cite{lu2023vehicle}. In these settings, applications often need selective access to past data rather than full sequential replay. Representative examples include city-scale mobility and infrastructure analytics from aggregated multi-vehicle sensing~\cite{xia2023automated}, predictive maintenance and fleet management based on long-horizon telemetry~\cite{theissler2021predictive}, and data-driven development workflows that search historical perception and driving data to surface rare events, edge cases, or distribution gaps~\cite{bolte2019towards,yin2017use}. Systems perspectives on autonomous vehicle data management similarly emphasize that many consumers behave like query workloads, requesting subsets of history defined by time windows, modalities, and sampling policies~\cite{kazhamiaka2021challenges}. The data-management community also extensively studied queryable mobility and spatio-temporal data for managing, indexing, and querying large trajectory collections~\cite{Eldawy2015SpatialHadoop,Ding2018UlTraMan,Li2020TrajMesa}. Thus, the problem is not only how to record streams, but how to retain compact temporal history that remains useful for later queries.

This demand for long-horizon AV data retention and selective historical access is amplified by the supply-side characteristics of AV data. Modern AV sensing is high volume and heterogeneous, and published measurements and surveys report that daily data generation can reach multiple terabytes depending on the sensor suite and driving duration \cite{wang2024quantitative}. Sustained ingestion at these rates can quickly exhaust onboard capacity, and it can also stress flash-based storage endurance under continuous write-intensive workloads, especially when write amplification and device endurance limits are considered \cite{snia_endurance_whitepaper}. At the same time, relying on continuous offload is often unrealistic: vehicular connectivity can be intermittent and bandwidth variable due to mobility and channel dynamics, which complicates high bandwidth data transfer in practice \cite{cheng2014vehicularoffloading}.

These vehicle computing applications and the scale of AV sensor data motivate an onboard data architecture that supports efficient sustained writes while enabling selective access to historical data, rather than treating storage primarily as a sequential log for offline replay.

\paragraph{Data Management Systems Gap Analysis}\label{bg:current-storage}


HydraSpace is one of the closest AV-oriented designs. It examines automotive logging under high-bandwidth heterogeneous sensors and proposes compression-centric techniques to reduce footprint~\cite{wang2020hydraspace}. However, this work focuses only on general compression without considering the downstream data usage, and does not consider the lower-level streaming write pressures.

The field-related robotics logging frameworks, such as ROS bag and MCAP, are widely used to support recording and offline replay. Recent work studying ROS/recording stacks reports limitations in query performance at scale and challenges around querying data concurrently with writing, depending on the chosen format and index layout~\cite{xu2024rosfs}. These frameworks are effective for faithful replay workflows, but they are not specifically designed around vehicle computing application data usage.

Beyond these domain-specific systems, broader IoT, mobility, and edge data-management systems provide useful building blocks for AVS. IoTDB and TsFile support high-ingest and compact time-series storage, while recent IoTDB work improves compressed-domain time-series queries~\cite{Wang2020ApacheIoTDB,Zhao2024ApacheTsFile,Tang2025CompressIoTDB}. NebulaStream manages streams across the sensor-edge-cloud continuum~\cite{Lepping2023NebulaStream}, and AdaEdge adaptively selects compression on resource-constrained edge devices~\cite{Liu2024AdaEdge}. These systems show the value of efficient ingest, compression, and edge-aware processing, but they do not target long-horizon in-vehicle retention of large heterogeneous vehicle payloads.

Edge storage systems emphasize locality, but many designs assume networked edge sites, replication, or cloud archival. SEND stores raw and processed data in network-edge repositories using labels to guide placement before cloud archival~\cite{nicolaescu2021store}, while FogStore provides geo-distributed key-value access on top of Cassandra with replication-based availability~\cite{gupta2018fogstore}. These assumptions are difficult inside a vehicle with fixed capacity, limited power, and strict onboard resource budgets.

Storage-engine, hybrid-storage, visual-data, and raw-data query systems provide useful design principles but do not directly match vehicles’ continuous, time-ordered ingest of large multimodal payloads with later selective random access. LSM-based systems optimize key-value write/query tradeoffs~\cite{Dayan2018Dostoevsky}, hStorage-DB exploits DBMS semantic information for hybrid storage~\cite{Luo2012hStorageDB}, VSS optimizes video storage for analytics~\cite{Haynes2021VSS}, and Slalom supports indexed query over raw files~\cite{Olma2017Slalom}.

Overall, existing data-management systems do not fully match the constraints of autonomous vehicles. AVS addresses this gap with a vehicle-specific architecture that sustains high-rate heterogeneous ingest, performs utility-aware reduction, retains application-usable history under fixed onboard budgets, and supports time- and modality-selective access directly on the vehicle.

\paragraph{Research Challenges}
\label{bg:rqs}
Motivated by this goal and the gaps above, constructing an efficient, queryable onboard data management system for AVs introduces three domain-specific design challenges that guide AVS.

RC1 (Ingest-time reduction). How can high-rate AV streams be reduced at ingest time to lower storage footprint while retaining information that is relevant to downstream applications?

RC2 (Sustained ingest and device efficiency). How to maintain sustained ingestion while protecting the device's lifespan under continuous logging?

RC3 (Queryable retention and retrieval). How should onboard storage be organized to support low-cost, long-horizon retention while supporting selective queries?


\begin{figure}[!ht]
\centering
\includegraphics[width=.8\linewidth]{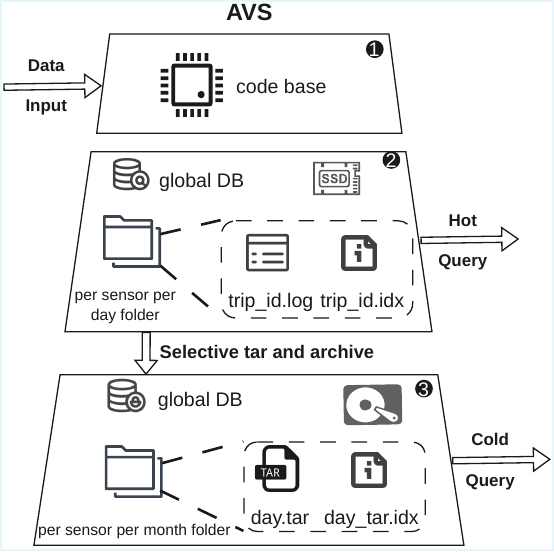}
\caption{The AVS architecture. 1. Code base for preprocessing. 2. SSD for recent temporal queries, with selective archival to 3. HDD, which also supports future cold queries.}
\label{fig:avs-system}
\vspace{-10pt}
\end{figure}
\section{AVS System Architecture}
\label{paper:design}
The overall AVS architecture is shown in Figure \ref{fig:avs-system}. AVS runs on a separate computing unit isolated from the AV main computer and receives sensor streams over Ethernet. Internally, AVS follows a single forward write path. Incoming messages first go through a lightweight preprocessing layer, then are committed to a hot SSD tier using append-only writes, and are later migrated to a cold HDD tier for long-term retention. Applications access both tiers through the same time and modality-aware query interface. Queries over recent time windows are served from SSD, while older windows are resolved from HDD archives using identical logical operators and indices. The responsibilities and operating principles of each AVS component are illustrated as follows.

\paragraph{Preprocessing Layer (RC1)}
AVS performs data reduction inside the ingest path rather than a separate offline job. The purpose of this stage is to reduce bytes before they reach persistent storage, while preserving the information needed by downstream workloads. Processing is modality aware and streaming oriented, meaning every message is handled within its publish period and either admitted in reduced form or dropped if it is redundant under the configured policy.

Concretely, the pipeline applies two classes of operators. First, redundancy removal removes repeated or low value samples from high rate streams using temporal or spatial criteria appropriate to each modality. Second, modality specific compression reduces the remaining payload size while maintaining task relevant fidelity. This stage produces a normalized temporal record stream that is ready for logging, and it defines the boundary between compute and data management: everything before this point is in memory processing, and everything after this point is durability and query. The operator set and modular implementation are described in Section~\ref{sec:operators}.

\begin{figure}[!ht]
\centering
\includegraphics[width=.9\linewidth]{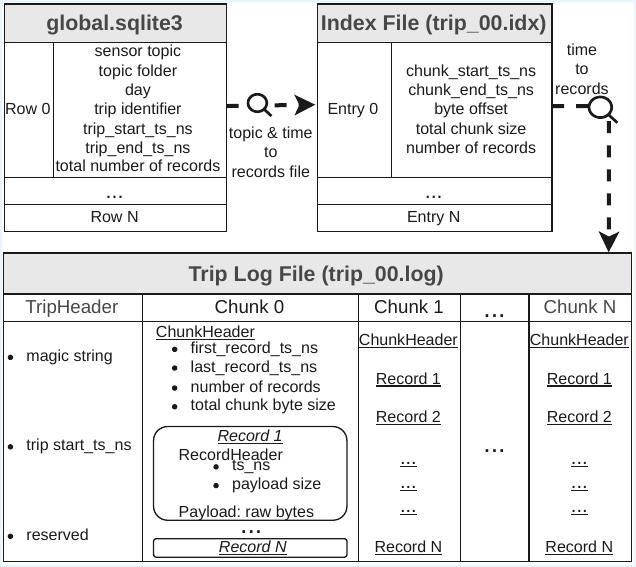}
\caption{Append-only logger structure.}
\label{fig:append-logger}
\vspace{-10pt}
\end{figure}

\paragraph{Append Oriented Logging and Query (RC2)}
Figure \ref{fig:append-logger} illustrates AVS’s unified append-oriented logging and query design. AVS records autonomous vehicle sensor data using strictly append-only writes to sustain high throughput ingestion under continuous workloads while minimizing write amplification and storage wear.

For each sensor topic on each day, AVS organizes data into trip-scoped streams. Each trip consists of a log file (trip\_XX.log) and a compact index file (trip\_XX.idx). The log file stores records sequentially. It starts with a fixed TripHeader that identifies the trip and its start time, followed by a sequence of chunks. Each chunk aggregates multiple records over a bounded time window or size threshold, and is written as a ChunkHeader followed by contiguous record payloads. This chunked layout amortizes synchronization cost across many records and preserves temporal locality for later reads.

Durability is defined at chunk granularity to tolerate power loss without complex recovery. After reboot, the logger scans the log sequentially, validates chunk boundaries, truncates any incomplete tail beyond the last valid chunk, and resumes appending to the same trip. This provides prefix durability with bounded loss limited to the most recent partial chunk.

To support efficient lookup, the per trip index stores one fixed size entry per chunk. Each entry records the chunk start and end timestamps, the byte offset in the log file, the chunk size, and the record count. Index entries are appended in time order, so queries can binary search the index to identify relevant chunks and then issue direct \texttt{pread} calls to fetch only the required byte ranges from the log.

At the system level, AVS maintains a global metadata database (global.sqlite3) that tracks trip-level descriptors across all sensors. Query execution follows three steps: it filters candidate trips using the global database, locates relevant chunks using the per-trip index, and retrieves only the referenced byte ranges from the log file. Because ingest and indexing are append-only, AVS can record and query concurrently without coordination that would require rewriting existing data, keeping ingest performance stable while providing time-bound access to both recent and historical data.

\begin{algorithm}[t]
\small
\caption{AVS SSD to HDD Archival (No Untar for cold query)}
\label{alg:avs_archive_short}
\KwIn{
SSD root $R_s$, HDD root $R_h$, cutoff $D_{cut}$
}
\KwOut{
Archived data on HDD and reclaimed SSD space
}

$C \leftarrow$ select archive candidates from SSD metadata;

\ForEach{$c \in C$}{
  $P_s \leftarrow$ SSD path of $c$\;
  $P_h \leftarrow$ HDD archive path of $c$\;
  Create $P_h$ if needed\;

  $F \leftarrow$ list trip log and index files under $P_s$\;
  Open temporary tar for write\;

  \ForEach{$f \in F$}{
    Write TAR header for $f$\;
    Record $(f,offset,size)$ in tar index\;
    Stream $f$ into tar\;
  }

  Finalize and \texttt{fsync} tar\;
  Write tar index file\;
  Atomically rename archive artifacts\;

  Copy metadata of $c$ from SSD DB to HDD DB\;
  Delete metadata of $c$ from SSD DB\;
  Remove $P_s$ from SSD\;
}
\end{algorithm}

\paragraph{Hierarchical Hot Cold Tier (RC3)}
AVS separates storage into a hot tier for low-latency access and a cold tier for cost-efficient retention. Newly ingested data are written to SSD, which serves recent queries with low read latency and high random access efficiency. This tier supports workloads that are most sensitive to time to first result.

As SSD capacity becomes constrained, AVS migrates older data to HDD to extend retention without requiring an oversized SSD. The prototype uses a concrete age- and capacity-based default: new data are first written to SSD by topic, day, and trip; once a topic day is complete, AVS archives it if it is older than the SSD retention window or if SSD free space falls below a threshold. In our experiments, the default policy keeps the most recent day on SSD and archives older completed topic days to HDD. Recency is a practical default because many vehicle-computing queries target recent data, such as post-drive inspection or abnormal-event analysis. The selector is still policy-driven: deployments can retune archival by topic, time range, or trip to avoid moving low-value records while preserving data likely to be queried later.

Archival from SSD to HDD is performed at the granularity of a sensor topic and day, as illustrated in Algorithm \ref{alg:avs_archive_short}. For each archive candidate, AVS packages all corresponding trip log and index files into a single tar archive. Crucially, this process preserves the append logger layout and does not require unpacking during future queries. Alongside the tar file, AVS generates a compact tar index that records the byte offset and size of each embedded file.

Metadata entries for archived data are migrated transactionally from the SSD-based global database to an HDD-resident database. AVS deletes the SSD copy only after the archive has been flushed, its tar index has been written, and the cold metadata are durable and queryable. This approach maximizes sequential I/O on HDD, minimizes filesystem fragmentation, and enables cold queries to locate and retrieve relevant data directly from the tar archive using indexed offsets.

\paragraph{Flexible Data Offload Mechanisms}
AVS provides two offload paths that match how AV data are consumed outside the vehicle. The first path targets bandwidth-constrained scenarios and supports selective network offload over Transmission Control Protocol (TCP). Rather than transferring full raw datasets, AVS can stream only the critical records, such as event-triggered frames plus their timestamps and minimal context, enabling near real-time sharing with a backend service while keeping transfer volume bounded.

The second path targets bulk export for offline analysis and supports physical offload by removing the HDD cold tier. Archived data are stored as self-contained tar files with associated tar indices and metadata databases, so the HDD can be mounted on an external machine and queried or copied immediately without additional preprocessing. This enables high-throughput transfer to data centers or partners when network offload would be impractical.

\section{Modular Mechanisms}
\label{paper:implementation}
Building on the AVS architecture, this section instantiates the data-management mechanisms that make long-horizon in-vehicle history practical: ingest-time preprocessing operators and lower-layer storage substrates. The preprocessing layer decides which sensor records should enter storage through modality-aware reduction and compression, while the storage substrate decides how those records are stored and accessed. This separation reflects a central observation for AV data systems: reducing bytes is not sufficient unless the retained records remain useful, and retaining useful records is not sufficient unless the physical layout supports sustained ingest and selective access. We therefore study both layers as modular design choices in the AVS pipeline before evaluating the complete prototype on real vehicle traces.

\subsection{Preprocessing Operators}
\label{sec:operators}
Persisting all raw AV data is impractical because sensor streams are high-rate, redundant, and application-dependent. AVS therefore treats preprocessing as an ingest-time data-management decision, rather than a generic offline compression step. Before data reaches persistent storage, AVS selects modality-specific reduction and compression settings that fit within the sensor period, preserve downstream utility, and reduce hot-tier write pressure.

We use the KITTI dataset~\cite{Geiger2013IJRR} as a controlled calibration workload. KITTI provides synchronized camera and LiDAR streams with standard odometry and tracking tasks, allowing AVS to evaluate each preprocessing choice by both footprint reduction and retained task utility. These tasks serve as retention-sensitive probes, not the only target applications of AVS. The selected parameters are conservative prototype defaults, and the same AVS knobs can be retuned for different sensors, domains, or application requirements. Thus, this section demonstrates a utility-aware retention workflow for AV data, rather than prescribing universal thresholds.

\subsubsection{LiDAR Downsampling and Compression}
\label{sec:lidar_ops}
\paragraph{Downsampling.}

Modern LiDAR sensors produce dense point clouds at 10--20\,Hz. Raw point density is often redundant for downstream tasks that depend on geometric structure rather than per point detail. Excess points increase storage and compute without commensurate gains. Among real-time preprocessing methods~\cite{9766181}, voxel grid downsampling is widely adopted due to its simplicity, streaming friendliness, and predictable cost~\cite{5980567}. It partitions 3D space into voxels of edge length $r$ and retains one representative point per voxel, typically the centroid. For a point cloud $\mathcal{P}=\{\mathbf{p}_1,\dots,\mathbf{p}_N\}$, the representative point for voxel $v$ is
\begin{equation}
\mathbf{p}_v = \frac{1}{|v|}\sum_{\mathbf{p}_i \in v}\mathbf{p}_i .
\end{equation}


The voxel size $r$ must balance storage savings against the accuracy loss that can occur when geometric detail is removed. To quantify this tradeoff, reduction utility is evaluated on 11 sequences of KITTI odometry with KISS-ICP~\cite{vizzo2023kiss} as a representative LiDAR data usage. KISS-ICP is robust and avoids manual tuning, which is a common proxy for localization and mapping consumers. Raw scans are first processed to establish a baseline, then voxel sizes are swept from 0.1 m to 1.0 m. Evaluation reports both trajectory quality and system relevant costs. Trajectory quality is measured using Absolute Trajectory Error (ATE), defined as the Root Mean Square Error (RMSE) between estimated and ground truth translation, and Average Rotation Error (ARE), defined as angular deviation per meter. System costs are captured by per-frame output size and preprocessing latency.

\begin{figure}[H]
    \centering
    \includegraphics[width=1\linewidth]{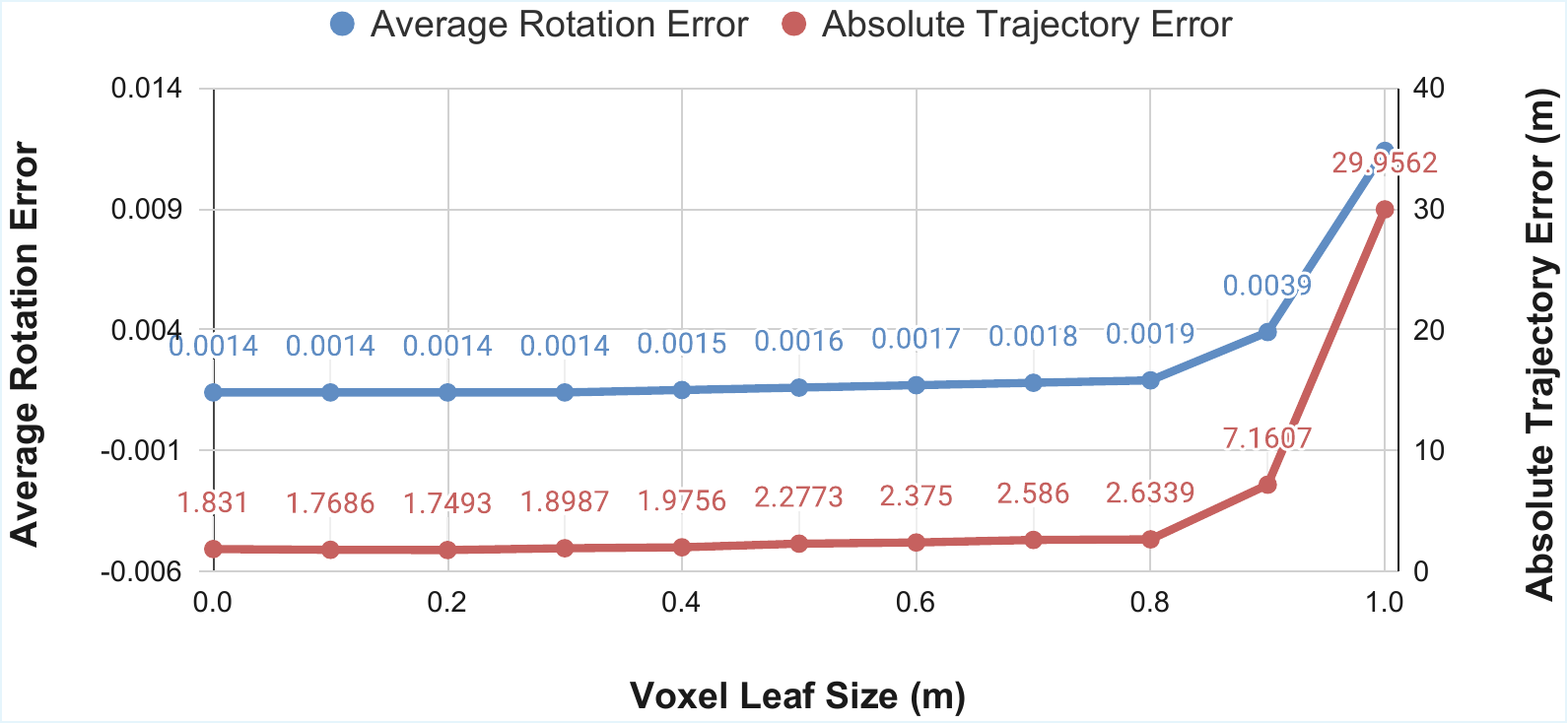}
    \caption{KISS-ICP performance. Red line: Absolute Trajectory Error (baseline is 1.831 m). Blue line: Average Rotation Error (baseline is 0.0014 deg/m)}
    \label{fig:kiss-icp-performance}
    \vspace{-10pt}
\end{figure}

\begin{figure}[H]
    \centering
    \includegraphics[width=1\linewidth]{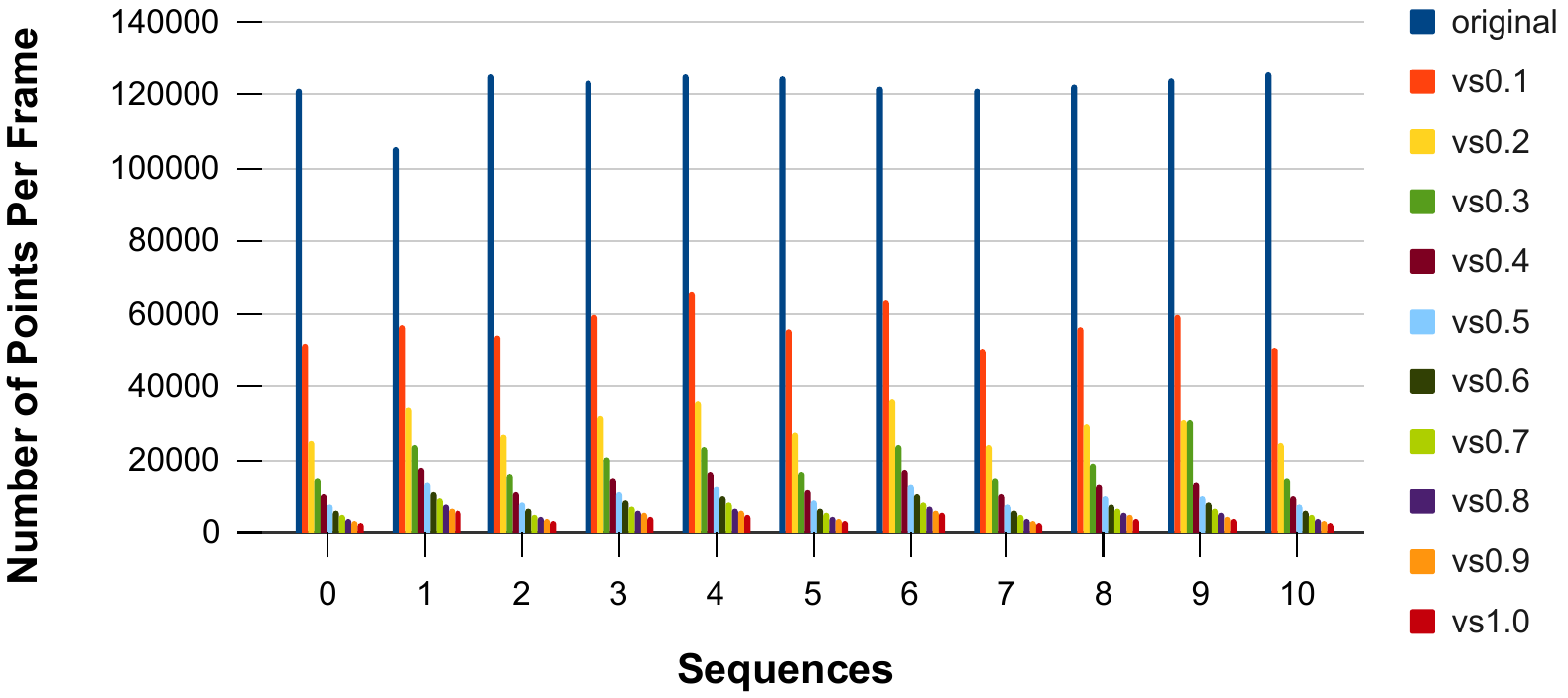}
    \caption{Number of points per frame comparison for different voxel leaf sizes and sequences. (Larger voxel leaf size, more aggressive data reduction)}
    \label{fig:voxel-vs-points}
    \vspace{-10pt}
\end{figure}

Figures~\ref{fig:kiss-icp-performance} and~\ref{fig:voxel-vs-points} show that larger voxels monotonically reduce points and file size. Odometry quality remains flat up to 0.3\,m and begins to degrade beyond 0.4\,m; at 1.0\,m the pipeline fails. Table~\ref{tab:voxel-vs-file-size} reports the detailed size and latency impact for the 0.2\,m setting. We select $r=0.2$\,m because it provides the best accuracy size tradeoff: on average, storage falls to 24.38\% of the original while ATE and ARE match the raw baseline, with approximately 9\,ms preprocessing overhead per scan, while within a 10\,Hz budget.

\begin{table}[t]
\centering
\caption{Comparison of LiDAR sequences before and after 0.2\,m filtering.}
\label{tab:voxel-vs-file-size}
\resizebox{\columnwidth}{!}{%
\begin{tabular}{*{8}{c}}
\toprule
\multirow{2}{*}{\textbf{Seq.}} &
\multicolumn{2}{c}{\textbf{\# Points / Frame}} &
\multicolumn{2}{c}{\textbf{Folder Size (MB)}} &
\multirow{2}{*}{\textbf{Ratio}} &
\multirow{2}{*}{\textbf{File Size Keep (\%)}} &
\multirow{2}{*}{\textbf{Latency (ms)}} \\
\cmidrule(lr){2-3}\cmidrule(lr){4-5}
& \textbf{Before} & \textbf{Filtered} & \textbf{Before} & \textbf{Filtered} & & & \\
\midrule
0  & 121494 & 25155 & 8800 & 1800 & 4.83 & 20.45\% & 9.18 \\
1  & 105682 & 34543 & 1900 & 609  & 3.06 & 32.03\% & 8.48 \\
2  & 125620 & 26663 & 9400 & 2000 & 4.71 & 21.28\% & 9.48 \\
3  & 123971 & 32277 & 1600 & 414  & 3.84 & 25.86\% & 9.74 \\
4  & 125718 & 36271 &  545 & 157  & 3.47 & 28.86\% & 10.17 \\
5  & 125038 & 27679 & 5500 & 1200 & 4.52 & 21.82\% & 9.62 \\
6  & 122300 & 36748 & 2200 & 647  & 3.33 & 29.42\% & 9.33 \\
7  & 121331 & 24297 & 2100 & 428  & 4.99 & 20.38\% & 9.23 \\
8  & 122594 & 29744 & 8000 & 1900 & 4.12 & 23.75\% & 9.64 \\
9  & 124384 & 31061 & 3200 & 791  & 4.00 & 24.71\% & 9.20 \\
10 & 125910 & 24557 & 2400 & 472  & 5.13 & 19.66\% & 8.70 \\
\midrule
\textbf{Avg.} & \textbf{122186} & \textbf{29909} & \textbf{4150} & \textbf{947} & \textbf{4.20} & \textbf{24.38\%} & \textbf{9.34} \\
\bottomrule
\end{tabular}%
}
\vspace{-5pt}
\end{table}
\paragraph{Compression.}
After downsampling, AVS applies a second step that targets byte-level compression before they are persisted on SSD. Two widely used families are considered. PCL Octree compression recursively partitions space and quantizes points at a chosen resolution~\cite{5980567}, providing a tunable tradeoff between size, fidelity, and runtime cost. In contrast, LASzip compresses LAS into LAZ using predictive coding and entropy coding designed around common LiDAR attributes, and is often used when high compression ratio and low reconstruction error are desired, at the cost of a more complex codec~\cite{isenburg2013laszip}.

\begin{figure}[t]
    \centering
    \includegraphics[width=1\linewidth]{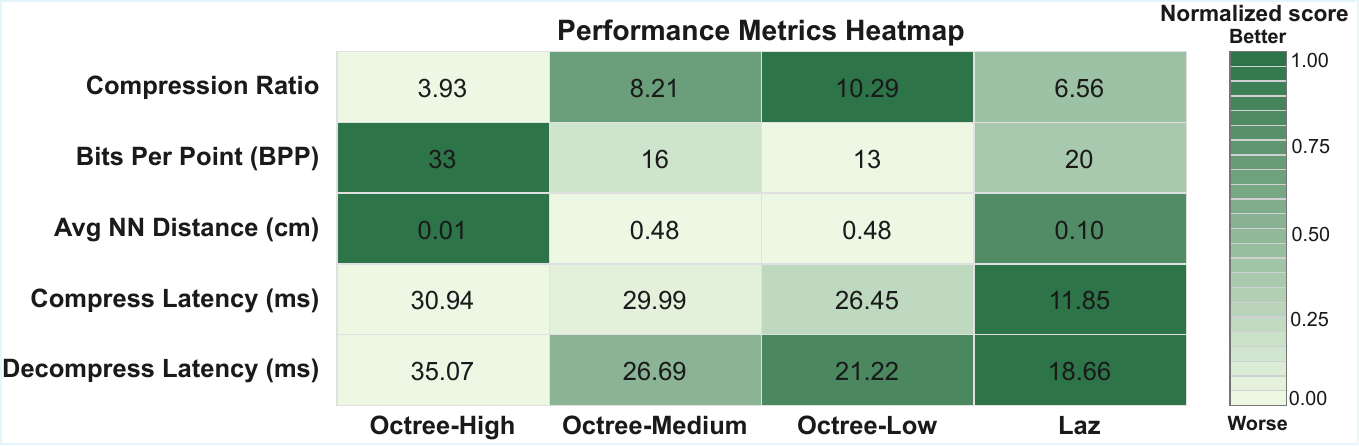}
    \caption{Lidar compression benchmark results. (Compression ratio and BPP larger is better, Avg NN Distance, compress and decompress latency, smaller is better)}
    \label{tab:lidar-compression-bench}
    \vspace{-10pt}
\end{figure}

Given these distinct tradeoffs, the choice is made empirically on the same 11 KITTI  sequences used for downsampling. Octree is evaluated across multiple resolutions to expose its fidelity size knob, and LAZ is evaluated as the representative LASzip based option. The benchmark reports four metrics that map directly to AVS goals: compression ratio (original BIN size divided by compressed size) and bits per point for hot tier footprint, average nearest neighbor distance after decompression for geometric distortion, and compression plus decompression latency for ingest and query overhead. Figure \ref{tab:lidar-compression-bench} summarizes the results. Lower and medium Octree resolutions can be competitive in compression ratio, but they introduce larger geometric error and higher encode decode cost than LAZ. Increasing Octree resolution further raises overhead without a clear improvement in reconstructed geometry for this workload. Overall, LAZ provides the most favorable balance of footprint reduction, geometric preservation, and runtime.

\begin{table}[t]
\centering
\caption{The KISS-ICP average performance over 11 KITTI LiDAR sequences: KITTI vs.\ VS0.2 vs.\ LAZ.}
\label{tab:laz-kiss-icp}
\begin{tabular}{cccc}
\toprule
\textbf{Metrics} & \textbf{KITTI} & \textbf{VS0.2} & \textbf{LAZ} \\
\midrule
Average Translation Error (\%) & 0.5164 & 0.5121 & 0.5136 \\
Average Rotation Error (deg/m) & 0.0014 & 0.0014 & 0.0014 \\
Absolute Trajectory Error (m)  & 1.8310  & 1.7493 & 1.7486 \\
Absolute Rotation Error (rad)  & 0.0801 & 0.0582 & 0.0571 \\
Average Frequency (Hz)          & 85.2727 & 94.3636 & 106.9091 \\
Average Runtime (ms)            & 12.6364 & 11.7273 & 10.6364 \\
\bottomrule
\end{tabular}%
\vspace{-10pt}
\end{table}

To validate downstream impact, we run KISS-ICP on three inputs per sequence: original point clouds, voxel downsampled clouds ($r=0.2$\,m), and LAZ compressed then decompressed clouds after the same downsampling. Table~\ref{tab:laz-kiss-icp} shows LAZ performance closely matches the downsampled baseline. So AVS adopts voxel downsampling with r equal to 0.2 m followed by LAZ compression as the LiDAR operator stack.

\subsubsection{Image Deduplication and Compression}
\label{sec:image_ops}

\paragraph{Deduplication.}
Camera streams are typically at a higher rate than LiDAR and exhibit strong temporal redundancy, especially during low-motion driving such as congestion and parking. Logging every frame wastes capacity and inflates write pressure on the hot tier. AVS removes redundancy using perceptual hashing (pHash)~\cite{zauner2010implementation}. Unlike cryptographic hashes, pHash captures low frequency structure and is robust to minor illumination and small viewpoint changes. We compute a 64-bit signature by applying Discrete Cosine Transform (DCT) to a resized grayscale image, extracting low frequency coefficients, and thresholding by their mean:
\begin{equation}
h_i =
\begin{cases}
1, & \text{if } c_i \ge \mu \\
0, & \text{if } c_i < \mu
\end{cases}
\end{equation}
where $c_i$ is a low frequency DCT coefficient and $\mu$ is the mean. Similarity is measured by Hamming distance:
\begin{equation}
\text{Hamming}(H_1,H_2)=\sum_{i=1}^{64}\mathbf{1}(H_{1,i}\neq H_{2,i}).
\end{equation}
If the distance is below a threshold $\tau$, the new frame is treated as redundant and discarded.

\begin{figure}[h]
  \centering
  \includegraphics[width=\linewidth]{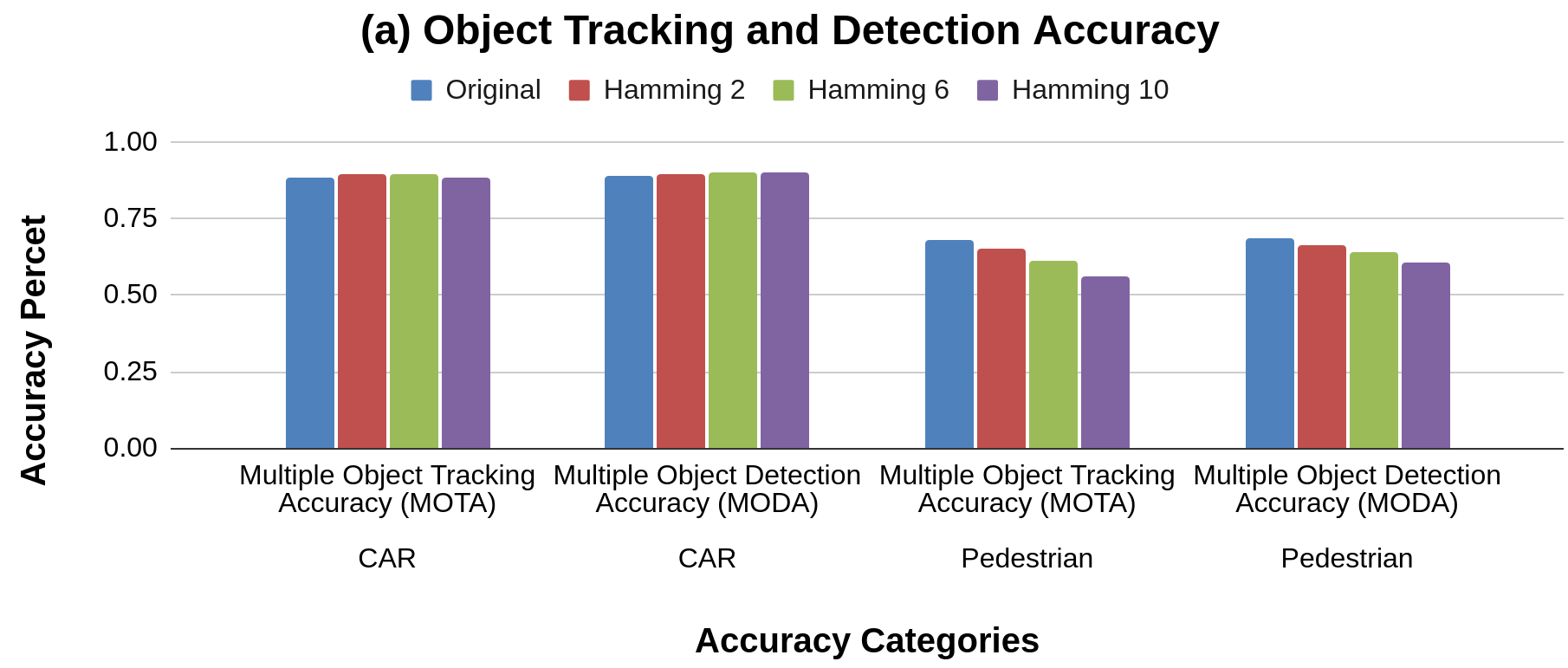}\par\medskip
  \includegraphics[width=\linewidth]{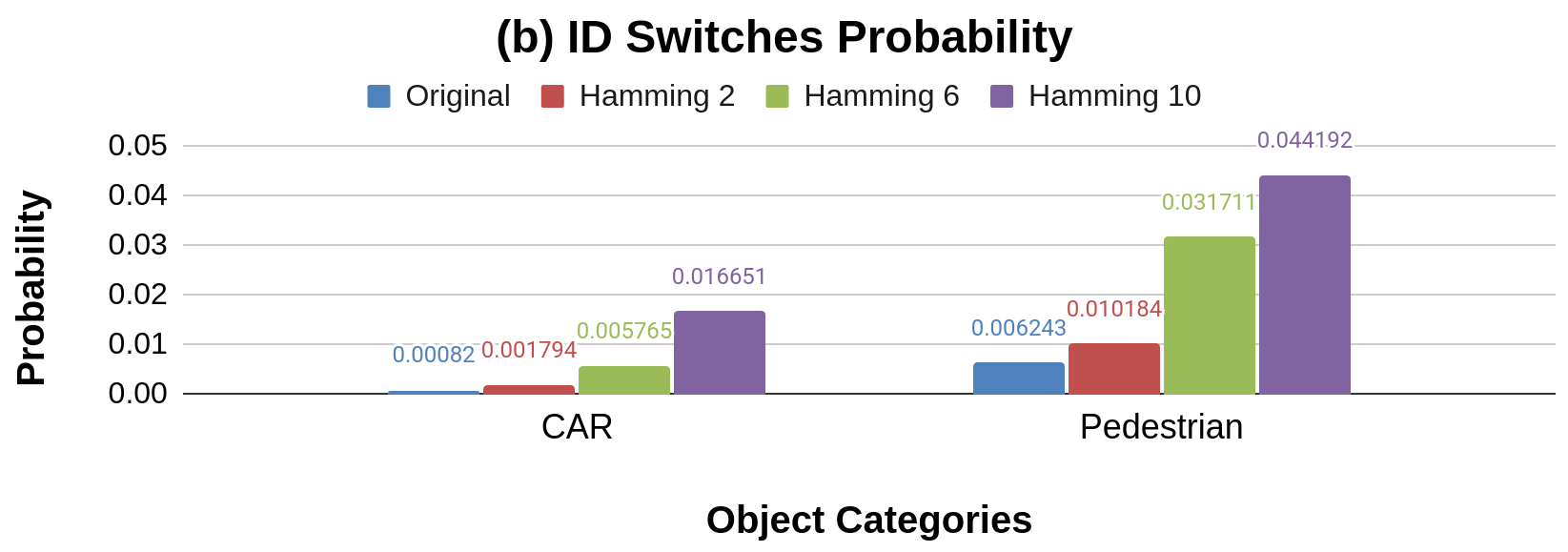}
  \caption{CenterTrack performance with different deduplication settings. (a) Shows multiple object tracking and detection accuracy for cars and pedestrians (the larger the better). (b) Shows ID switches probability for cars and pedestrians (the smaller the better)}
  \label{fig:center-track-performance}
  \vspace{-10pt}
\end{figure}

\begin{figure}[h]
    \centering
    \includegraphics[width=1\linewidth]{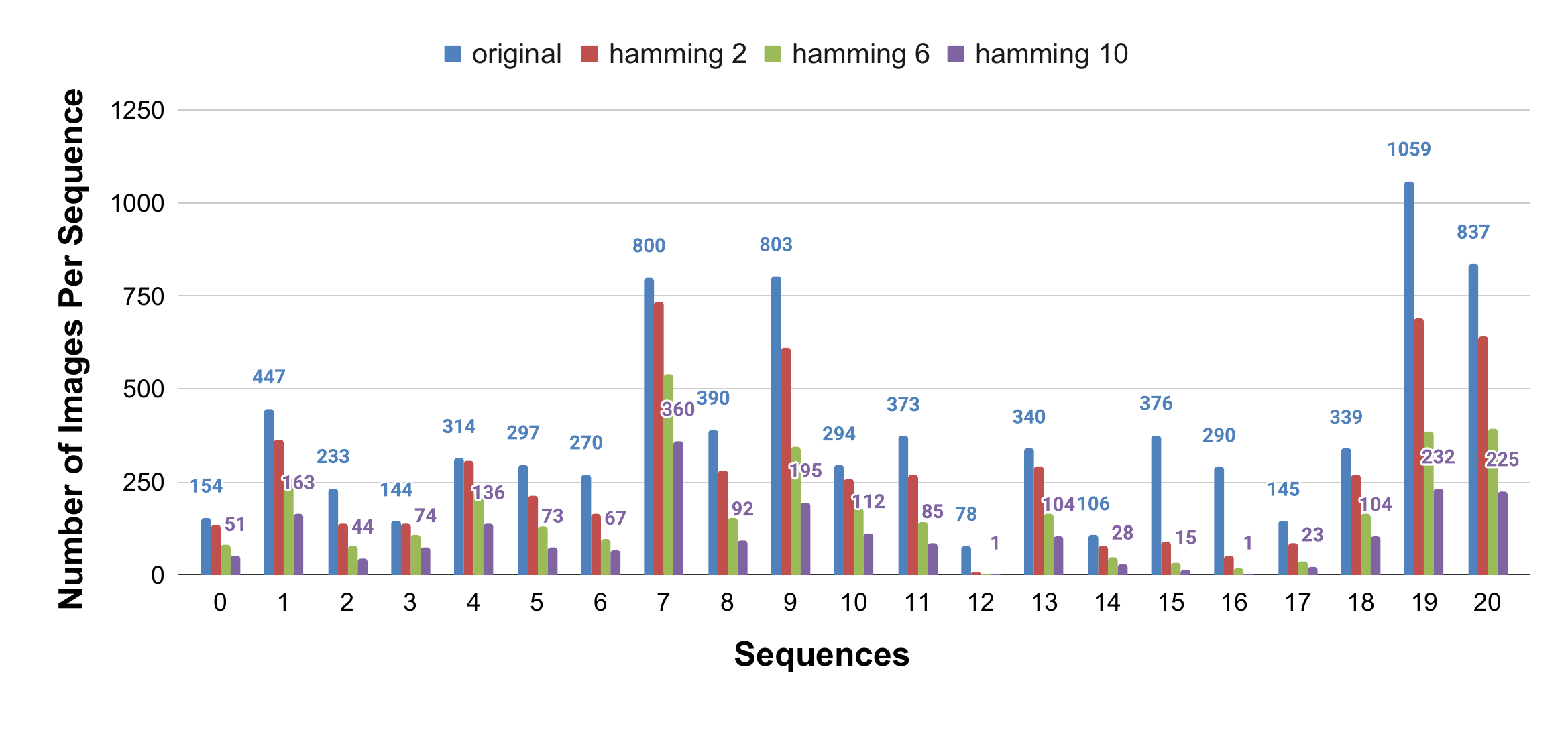}
    \caption{Number of images per sequence under different hamming distance settings.}
    \label{fig:phash-dedup}
    \vspace{-10pt}
\end{figure}

The threshold $\tau$ controls how aggressively frames are removed, so it is selected based on downstream impact rather than only on saved bytes. Evaluation is performed on KITTI Tracking (21 sequences at 10 Hz) using CenterTrack~\cite{zhou2020tracking} as a representative detection and tracking workload that benefits from adjacent frame cues. A baseline is established on the original image streams, then the same workload is run after deduplication with $\tau$ in {2, 6, 10}. Reported metrics include Multiple Object Tracking Accuracy (MOTA), Multiple Object Detection Accuracy (MODA), ID switches, retained frame ratio, and per-frame deduplication latency. Figure~\ref{fig:center-track-performance} shows that car tracking remains largely invariant to pruning, while pedestrian performance degrades as pruning becomes aggressive, reflected in increased ID switches. Figure~\ref{fig:phash-dedup} shows that $\tau=2$ yields 28.25\% reduction with about 2.2\,ms overhead per frame. So we adopt $\tau=2$ as a conservative operating point that yields significant savings with minimal utility loss.

\begin{table}[t]
\centering
\caption{Comparison of image compression. (Reported numbers are averaged across 21 KITTI image sequences.)}
\label{tab:image-format-compression}
\resizebox{\columnwidth}{!}{%
\begin{tabular}{ccccc}
\toprule
\textbf{Metrics} & \textbf{JPEG (85)} & \textbf{JPEG (95)} & \textbf{JPEG2000} & \textbf{WebP} \\
\midrule
Compression Time (ms) & 1.19 & 1.45 & 95.39 & 34.61 \\
Compression Ratio & 7.70 & 4.06 & 1.12 & 6.83 \\
\bottomrule
\end{tabular}%
}
\vspace{-10pt}
\end{table}

\begin{table}[t]
\centering
\caption{CenterTrack performance for car and pedestrian under hamming 2 deduplication with different compression settings. (Baseline is hamming 2 deduplication results.)}
\label{tab:image-format-center-track}
\resizebox{\columnwidth}{!}{%
\begin{tabular}{ccccc}
\toprule
\textbf{Object} & \textbf{Setting} & \textbf{MOTA} & \textbf{MODA} & \textbf{ID-switches} \\
\midrule
\multirow{3}{*}{Car} 
  & Hamming 2             & 0.8935 & 0.8953 & 0.0018 \\
  & Hamming 2 + JPG 85    & 0.8909 & 0.8935 & 0.0026 \\
  & Hamming 2 + JPG 95    & 0.8952 & 0.8972 & 0.0020 \\
\midrule
\multirow{3}{*}{Pedestrian} 
  & Hamming 2             & 0.6506 & 0.6608 & 0.0102 \\
  & Hamming 2 + JPG 85    & 0.6463 & 0.6600 & 0.0137 \\
  & Hamming 2 + JPG 95    & 0.6533 & 0.6635 & 0.0102 \\
\bottomrule
\end{tabular}%
}
\vspace{-10pt}
\end{table}
\paragraph{Compression.}
After deduplication, the remaining frames are further encoded before being written to the SSD to reduce the hot-tier footprint. AVS adopts per-image encoding instead of inter-frame codecs such as H.264, which rely on temporally predictive I/P/B frames organized in Group of Pictures (GOP) structures that couple adjacent frames and restrict independent random access~\cite{wiegand2003overview}. We evaluate JPEG, JPEG2000, and WebP, which offer different latency and compression tradeoffs. JPEG uses block DCT with quantization~\cite{wallace1991jpeg}, JPEG2000 applies a wavelet transform~\cite{taubman2000high}, and WebP uses intra prediction with entropy coding~\cite{google_webp_2010}.

Compression formats are compared on the KITTI image sequences using metrics that map directly to AVS constraints: encoding latency for ingest feasibility, compression ratio for footprint reduction, and CenterTrack performance to check downstream impact on the deduplicated streams. Table~\ref{tab:image-format-compression} shows that JPEG provides the lowest latency by a wide margin; JPEG2000 and WebP are substantially slower in our setting. Table~\ref{tab:image-format-center-track} shows JPEG quality 95 preserves tracking accuracy while still providing 4.06$\times$ size reduction with 1.45\,ms per image encoding time. Based on this tradeoff, AVS uses JPEG quality 95 as the image format.

\subsection{Lower Layer File System Selection}
\label{sec:lower_system}

Operators reduce bytes, but AVS feasibility also depends on the behavior of the storage substrate that implements durability and layout. AVS places recent data on SSD for low-latency access and migrates older data to HDD for long retention. The hot tier, therefore, stresses durable writes and metadata operations, while the cold tier stresses high throughput sequential I/O over large archives with minimal fragmentation.

We benchmark EXT4 and XFS due to their widespread deployment and contrasting allocation and metadata strategies~\cite{mohan2017analyzing,jaffer2019evaluating}. EXT4 is a stable general-purpose journaling filesystem with delayed allocation and extents. XFS is designed for high throughput and scalable allocation, often outperforming under large file and sequential workloads.


\begin{figure}[ht]
    \centering
    \includegraphics[width=1\linewidth]{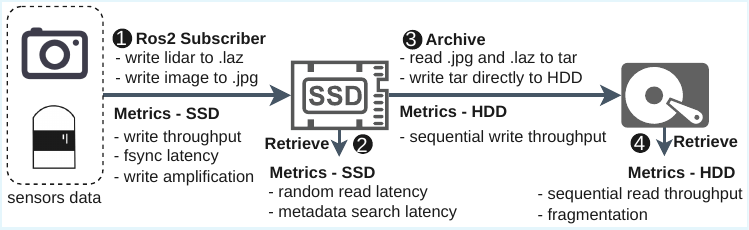}
    \caption{File system benchmark experiment logic.}
    \label{fig:file-system-bench}
\end{figure}

The benchmark follows the AVS data path and is illustrated in Figure~\ref{fig:file-system-bench}. LAZ and JPEG outputs are logged to SSD while measuring write throughput at 10Hz data logging, fsync latency, and write amplification. Logged data are then packed into tar archives and written to HDD to measure archival throughput. Retrieval like access is exercised by measuring small random read latency on SSD and metadata lookup overhead. Finally, tar archives are read back from HDD to measure sequential read throughput and fragmentation. We quantify fragmentation as
\begin{equation}
frag\_index = 1 - \left(\frac{largest\_extent\_bytes}{total\_file\_size\_bytes}\right)
\end{equation}
which captures whether large archives are placed contiguously.

\begin{table}[t]
\centering
\caption{SSD filesystem benchmark results: EXT4 vs.\ XFS. (Reported numbers are averaged across 11 LiDAR sequences and 21 image sequences.)}
\label{tab:SSD-filesystem-benchmark}
\begin{tabular}{@{}c c c c c@{}}
\toprule
\multirow{2}{*}{\textbf{Metrics}} & \multicolumn{2}{c}{\textbf{EXT4}} & \multicolumn{2}{c}{\textbf{XFS}} \\
\cmidrule(lr){2-3}\cmidrule(l){4-5}
 & \textbf{.jpg} & \textbf{.laz} & \textbf{.jpg} & \textbf{.laz} \\
\midrule
Average Write Throughput (MB/s)     & 2.06 & 1.00 & 2.06 & 1.09 \\
Average \texttt{fsync} Latency (ms) & 2.07 & 1.74 & 1.94 & 1.57 \\
Average Write Amplification         & 1.20 & 1.38 & 1.06 & 1.12 \\
4\,KiB Random Read Latency (ms)     & 4.93 & 5.27 & 4.91 & 5.16 \\
Metadata Search Latency (ms)        & \multicolumn{2}{c}{0.21} & \multicolumn{2}{c}{0.20} \\
Tar file size (MB)                  & \multicolumn{2}{c}{3842.31} & \multicolumn{2}{c}{3842.31} \\
\bottomrule
\end{tabular}%
\end{table}

\begin{table}[t]
\centering
\caption{HDD filesystem benchmark results: EXT4 vs.\ XFS. (Reported numbers are averaged across 11 LiDAR sequences and 21 image sequences.)}
\label{tab:HDD-filesystem-benchmark}
\begin{tabular}{ccc}
\toprule
\textbf{Metrics} & \textbf{EXT4} & \textbf{XFS} \\
\midrule
Sequential Write Throughput (MB/s)
  & 103.00 & 107.49 \\
Sequential Read Throughput (MB/s)
  & 119.11 & 124.71 \\
Fragmentation (frag\_index) & 0.66 & 0.00 \\
\bottomrule
\end{tabular}%
\vspace{-10pt}
\end{table}

Tables~\ref{tab:SSD-filesystem-benchmark} and~\ref{tab:HDD-filesystem-benchmark} summarize the results. On SSD, EXT4 and XFS provide similar write throughput; XFS yields slightly lower fsync latency and lower write amplification, improving durability cost under frequent commits. On HDD, the separation is clear: XFS provides higher sequential throughput and exhibits zero measured fragmentation for tar archives, while EXT4 shows substantial fragmentation (frag\_index $\approx 0.66$), harming long sequential reads for cold replay and analytics. Since AVS archives unstructured data into large tar files, XFS aligns better with the cold-tier access pattern. We therefore adopt XFS in the prototype.


\subsection{Logger Backend Comparison}
\label{sec:db_compare}

AVS designed a custom append-only logger because vehicle data recording is dominated by continuous, time-ordered ingest of large sensor payloads. This workload differs from a conventional database workload: records are rarely updated after arrival, queries are primarily bounded by topic and time, and durability must be achieved with low write amplification under sustained streaming writes. Therefore, AVS stores payloads in sequential trip logs and keeps only compact time-range metadata in indexes, rather than inserting every sensor record into a general-purpose database engine.

\begin{table}[H]
\centering
\caption{Database backend comparison under the same preprocessed workload. Lower is better for all metrics. (Reported numbers are averaged values.)}
\label{tab:db-backend-compare}
\resizebox{\columnwidth}{!}{%
\begin{tabular}{@{}cccc@{}}
\toprule
\textbf{DB Backend} & \textbf{Insert Latency (ms)} & \textbf{Query Latency (ms)} & \textbf{Final Size (MB)} \\
\midrule
SQLite & 0.4276 & 0.6900 & 741.5352 \\
RocksDB & 0.5685 & 4.6622 & 739.2103 \\
Append-only & 0.1712 & 1.7879 & 739.1076 \\
\bottomrule
\end{tabular}%
}
\vspace{-5pt}
\end{table}

To evaluate this design, we compare the AVS append-only logger with SQLite~\cite{sqlite} and RocksDB~\cite{rocksdb} using the same preprocessed camera, LiDAR, and GPS workload, consisting of 7,745 records and 1,000 time-range queries. Table~\ref{tab:db-backend-compare} shows that the append-only logger achieves the lowest insert latency and smallest final footprint, while SQLite provides lower query latency. This result reflects the intended tradeoff: AVS prioritizes predictable ingest, sequential payload writes, and low storage overhead, while accepting slightly higher query latency that remains suitable for time-window retrieval. The comparison shows why the logger is not simply a replacement for a database, but a storage layout specialized for AV ingest and access patterns.

\begin{figure}[H]
    \centering
    \includegraphics[width=1\linewidth]{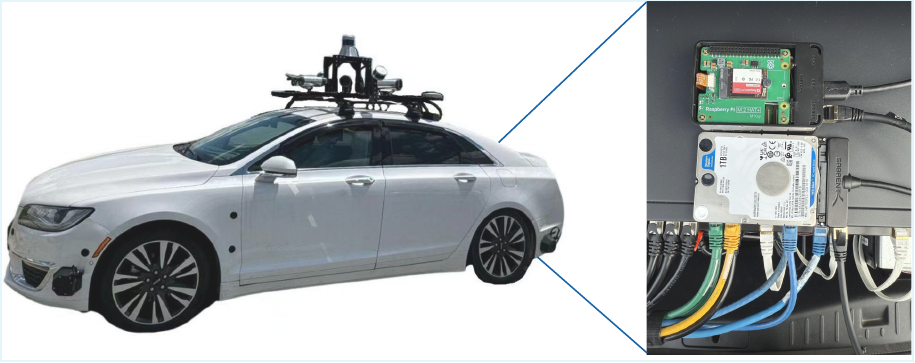}
    \caption{Lincoln MKZ based L4 autonomous driving platform with AVS prototype, connected through Ethernet.}
    \label{fig:hydraD}
\end{figure}
\section{AVS Prototype}
\label{paper:evaluation}
\begin{figure*}[ht]
    \centering
    \includegraphics[width=\linewidth]{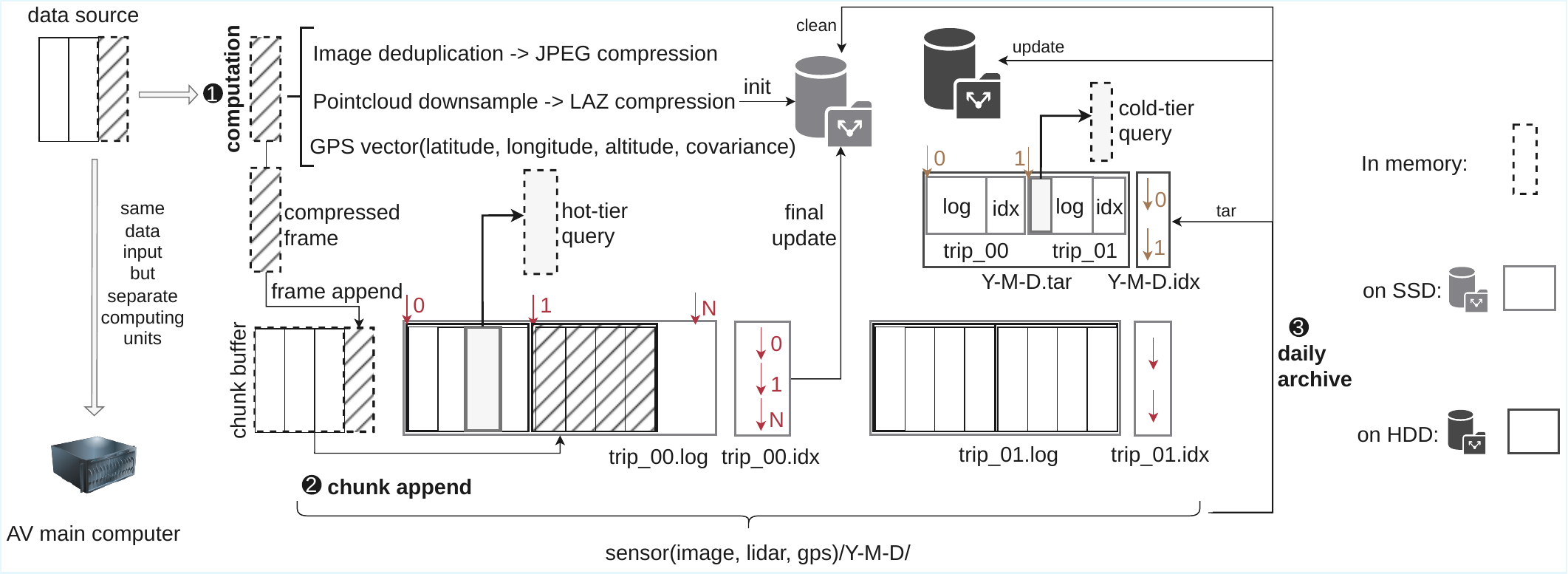}
    \caption{AVS prototype pipeline: (1) in memory reduction and compression, (2) chunked append logging to the SSD hot-tier with indices, (3) topic day archival into indexed tar files on the HDD cold tier, and SSD reclamation.}
    \label{fig:prototype-workflow}
    \vspace{-5pt}
\end{figure*} 
\subsection{Prototype Setup}
\label{sec:platform_data}
\paragraph{Platform.}
The AVS prototype is implemented on a Raspberry Pi 5 with a 256 GB NVMe SSD and a 1 TB WD10SPZX HDD, forming a practical two-tier in-vehicle storage hierarchy with only \$194 cost. The Raspberry Pi 5 integrates a Broadcom BCM2712 SoC with four Arm Cortex A76 cores at 2.4 GHz and 8 GB of RAM, which is sufficient to sustain concurrent processing. The NVMe SSD is partitioned into a system volume and a dedicated 100 GB hot tier that buffers recent data for real-time ingestion and low-latency access. The HDD serves as the cold archival tier for long-term retention and is attached through a USB to SATA bridge. Both tiers are formatted with XFS for consistent file system behavior across devices.

To avoid interfering with safety-critical workloads on the vehicle's main computer, AVS runs as an isolated computing device with its own OS stack, Ubuntu 24.04, and ROS2 Jazzy. It subscribes to sensor streams over Ethernet using ROS2 topics exported by the vehicle platform, enabling decoupled data collection without placing additional load on the primary compute. The overall hardware setup and system integration are illustrated in Figure~\ref{fig:hydraD}.


\paragraph{Data.}
We evaluate AVS using three days of mixed-time driving, with two trips and approximately 1 hour of driving per day. This duration is representative of typical U.S. daily driving time (around 60 minutes/day), so each ‘day’ in our traces reflects a realistic amount of on-road data generation for an individual vehicle~\cite{hungund_steinbach_2025_ads2024}. Data are collected on a Lincoln MKZ based Level~4 autonomous vehicle platform.

The workload consists of concurrent multi-modal sensor streams representative of practical AV deployments: one 10\,Hz Hesai Pandar64 LiDAR stream (\texttt{sensor\_msgs/PointCloud2}), one 10\,Hz front-facing Basler Ace mono8 camera images (\texttt{sensor\_msgs/Image}), and one 50\,Hz NovAtel OEM7 GNSS measurements (\texttt{gps\_msgs/GPSFix}). All sensors operate simultaneously throughout each trip, stressing AVS with realistic concurrent ingestion, processing, and storage demands.
\paragraph{Workflow.}
\begin{figure}[t]
    \centering
    \includegraphics[width=1\linewidth]{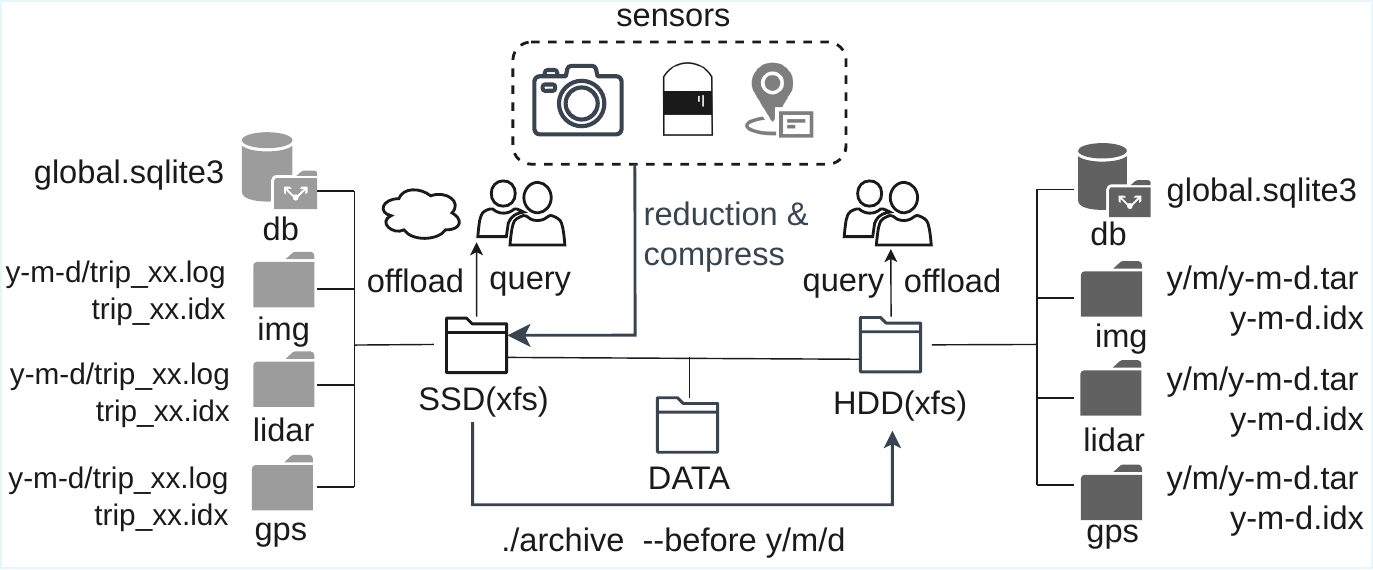}
    \caption{AVS prototype data folder structure overview.}
    \label{fig:proto-folder}
    \vspace{-10pt}
\end{figure}

The prototype detailed workflow is shown in Figure~\ref{fig:prototype-workflow}. The overall folder structure for different sensor data is shown in Figure~\ref{fig:proto-folder}. AVS records each sensor stream as a continuous sequence of frames that flow through a unified append-only pipeline. Raw camera, LiDAR, and GPS data are first processed in the computation stage, where images are deduplicated and JPEG compressed, point clouds are downsampled and LAZ compressed, and GPSFix messages are reduced to a compact trajectory oriented representation containing latitude, longitude, altitude, and position covariance, which is commonly needed for historical trajectory queries as in NavSatFix. Each processed frame is appended to an in-memory chunk buffer. When the buffer reaches 2 MB or 250 ms threshold, the chunk is sealed and atomically flushed to SSD as a contiguous segment in the same trip log file, while a corresponding index entry recording the time range, file offset, chunk size, and record count is appended to the trip index file. All writes are sequential. Each trip is defined from engine start to engine off. On SSD, data is organized by topic and day as topic/year\_month\_day/trip.log and trip.idx files, with a global sqlite3 database storing one row per trip containing topic folder, day, trip id, start time, end time and total record count. The detailed metadata information is the same as the design mentioned before in Figure~\ref{fig:append-logger}. 
\subsection{Performance on Real Driving Trace}
\label{sec:real_trace}

\begin{table*}[t]
\centering
\caption{Recording performance comparison: AVS vs. rosbag2 (sqlite and mcap, raw and zstd). AVS is evaluated as an end-to-end utility-aware retention pipeline; footprint savings mainly reflect preprocessing, while ingest, write, and resource metrics reflect the storage path. Notes: total system memory is 8 GB; total CPU capacity is 400\% (4 cores).}
\label{tab:prototype-recording-bench}
\resizebox{\linewidth}{!}{%
\begin{tabular}{*{9}{c}}
\toprule
\multicolumn{2}{c}{} &
\multicolumn{3}{c}{\textbf{Workload}} &
\multicolumn{3}{c}{\textbf{NVMe SMART (deltas)}} \\
\cmidrule(lr){3-5} \cmidrule(lr){6-8}
\textbf{Day} & \textbf{Method} & \textbf{Data Size (GB)} &
\textbf{CPU avg/max (\%)} & \textbf{RSS avg/max (MB)} &
\textbf{Data Units Written (GB)} & \textbf{WAF} & \textbf{Host Write Commands} \\
\midrule
\multirow{5}{*}{Day 1 (4104s)}
& AVS                & \textbf{23.44}  & 82.86 / 176.40 & 175.86 / 179.55 & \textbf{24.57} & \textbf{1.048} & \textbf{22{,}929} \\
& rosbag2 sqlite     & 201.56 & 27.06 / 100.10  & 82.88 / 96.08   & 211.36 & 1.049 & 198{,}827 \\
& rosbag2 sqlite zstd& 117.52 & 50.89 / 180.90 & 96.92 / 122.30   & 123.81 & 1.054 & 118{,}750 \\
& rosbag2 mcap       & 201.21 & \textbf{14.76} / 180.40 & \textbf{81.32} / 98.18   & 211.51 & 1.051 & 167{,}904 \\
& rosbag2 mcap zstd  & 116.69 & 43.23 / 200.50 & 111.42 / 121.71  & 122.31 & 1.048 & 99{,}329 \\
\midrule
\multirow{5}{*}{Day 2 (4173s)}
& AVS                & \textbf{25.42}  & 87.73 / 113.50 & 178.48 / 181.00 & \textbf{26.64} & \textbf{1.048} & \textbf{24{,}283} \\
& rosbag2 sqlite     & 204.08 & 28.77 / 199.30  & \textbf{89.01} / 109.50   & 214.14 & 1.049 & 183{,}190 \\
& rosbag2 sqlite zstd& 121.46 & 50.95 / 173.50 & 106.10 / 122.26 & 127.41 & 1.050 & 111{,}658 \\
& rosbag2 mcap       & 201.29 & \textbf{15.20} / 170.90  & 89.83 / 101.91   & 211.09 & 1.050 & 165{,}286 \\
& rosbag2 mcap zstd  & 121.59 & 42.78 / 199.70 & 112.92 / 126.14 & 127.58 & 1.051 & 111{,}240 \\
\midrule
\multirow{5}{*}{Day 3 (3957s)} 
& AVS                & \textbf{24.21}  & 87.13 / 135.70 & 178.06 / 184.35 & \textbf{25.29} & \textbf{1.045} & \textbf{22{,}713} \\
& rosbag2 sqlite     & 191.79 & 28.94 / 197.00 & 91.77 / 106.85   & 200.92 & 1.048 & 175{,}086 \\
& rosbag2 sqlite zstd& 112.64 & 52.21 / 194.00 & 100.12 / 122.89 & 118.12 & 1.049 & 104{,}189 \\
& rosbag2 mcap       & 194.17 & \textbf{15.71} / 195.50  & \textbf{85.45} / 99.17  & 203.68 & 1.049 & 160{,}122 \\
& rosbag2 mcap zstd  & 112.63 & 43.52 / 194.20 & 107.83 / 123.00 & 118.23 & 1.050 & 93{,}916 \\
\bottomrule
\end{tabular}%
}
\vspace{-10pt}
\end{table*}

\paragraph{Stream Ingest Performance and Stability.}
We evaluate ingest overhead on real driving traces by recording the same set of camera, LiDAR, and GPS topics to the NVMe hot tier using AVS and rosbag2. For rosbag2, we test both backends, sqlite3 and mcap, and for each backend we test raw logging and zstd message level compression. All runs use the same trip duration and the same storage target. We attribute compute overhead by periodically sampling CPU utilization and resident set size and aggregating over the full process trees of the recorder and its spawned executables, so that all ingestion-related work is counted consistently. We quantify storage cost using NVMe SMART deltas collected immediately before and after each run, reporting net data units written, write amplification factor, and host write command count.

\begin{figure}[ht]
    \centering
    \includegraphics[width=0.9\linewidth]{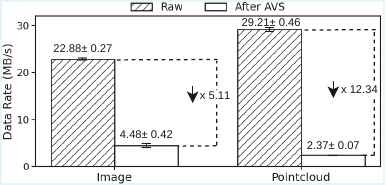}
    \caption{Modality effective data rate after AVS computation operators during ingest. Numbers are reported as mean $\pm$ 95\% Confidence Interval (CI95).}
    \label{fig:reduction-performance}
    \vspace{-10pt}
\end{figure}

Table~\ref{tab:prototype-recording-bench} shows three consistent trends. First, AVS substantially reduces the final recorded footprint compared with replay-oriented bagging, storing 8.0 to 8.7$\times$ fewer bytes than raw rosbag2 and 4.5 to 5.0$\times$ fewer bytes than rosbag2 with zstd. These savings come mainly from AVS’s modality-aware reduction and compression for images and point clouds, as summarized in Figure~\ref{fig:reduction-performance}. The figure reports reduction on raw sensor payloads only, whereas the table reports the final on disk trip size, which also includes metadata and logging layout overhead.

Second, AVS reduces SSD write traffic at the command level, which directly affects device overhead. With the write amplification factor staying close to 1.048 for all methods, AVS consistently issues far fewer host write commands. These indicate larger, more sequential writes, which is consistent with AVS append-oriented chunking and reduces per-write firmware work on the NVMe. Net data units written also track the stored data size closely for AVS, for example 24.57 GB written for 23.44 GB stored, indicating minimal additional internal write traffic beyond the user visible output.

Third, AVS shifts overhead from storage to compute, but remains stable under continuous ingest. AVS shows higher average CPU utilization than rosbag2, because it performs data reduction and compression online. However, its peak CPU remains comparable to rosbag2 peaks in the same traces, and does not exhibit runaway growth across longer runs. Memory use is also stable, staying around 176 to 184 MB across trips, with a very small max-min average gap, suggesting bounded buffering and no accumulating backlog during ingest. Although rosbag2 uses less memory in these runs, the AVS footprint remains a small fraction of the available 8 GB system memory, leaving headroom for additional operators.
\paragraph{Power Loss and Recovery.}
AVS provides prefix durable recovery for append-only trip logs. Each sensor stream is recorded as a sequence of fixed-size chunks appended sequentially to disk, with durability defined at chunk granularity. To sustain high throughput ingestion and reduce write amplification, AVS appends data into the OS page cache and relies on the kernel to schedule physical SSD writes. Consequently, data buffered but not yet flushed at the time of a power failure forms a small in-flight tail that may be lost.
After an unclean shutdown, AVS deterministically recovers by reopening the active trip and reconstructing the last durable prefix. The logger identifies the last fully persisted chunk boundary by scanning chunk headers, truncates any unsafe tail beyond this boundary, repairs global SQLite metadata and the corresponding index file, and resumes appending to the same trip. All data before the boundary is preserved, while only the in-flight tail is discarded.

\begin{figure}[ht]
    \centering
    \includegraphics[width=1\linewidth]{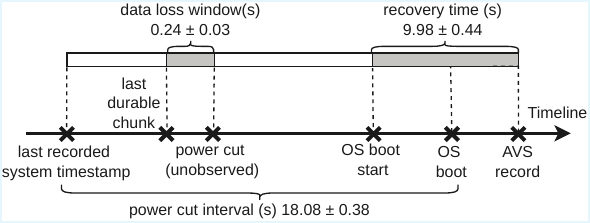}
    \caption{Power loss and recovery behavior of AVS (Reported numbers are averaged over 10 runs shown as mean $\pm$ CI95).}
    \label{fig:power-loss}
\end{figure}


Durability is evaluated using controlled power loss experiments that forcibly reboot the system during active ingestion. Three metrics are reported. Power cut interval is the elapsed time from the last pre-crash recorded system timestamp to the OS boot. Recovery time is the time from boot start until AVS restores a consistent log state and resumes operation. Data loss window is the interval between the timestamp of the last durable chunk and the unobserved power cut, which bounds the lost suffix. As shown in Figure~\ref{fig:power-loss}, the measured reboot gap is 18.08 s on average across 10 runs. AVS completes recovery in 9.98 s and limits the data loss window to 0.24 s. No cyclic redundancy check (CRC) mismatches are observed, confirming that the recovered prefix remains bit identical while power failures result only in bounded suffix loss.
\paragraph{Temporal Query Behavior.}
AVS supports retrieval by topic and time window, returning records in increasing timestamp order. A query consults the global trip catalog to find trips containing the topic, scans each selected trip’s per-trip index to locate chunks overlapping the time window, and then streams matching records by parsing chunk and record headers.

The same query path applies to archived trips, closed trips on SSD, and the currently recording open trip, but the visibility boundary differs by state. For hot closed trips, the log is finalized and fully indexed, so queries scan an immutable SSD-resident log. For archived trips, the catalog and indices still narrow the search, but chunk retrieval incurs archive/HDD access before records can be streamed. For an open trip, AVS supports querying while ingestion continues: records become queryable only after their chunk is flushed to the SSD log and its index entry is durable; records still buffered in the active chunk are not visible. Since most third-party applications primarily query historical data rather than a live sensor stream, AVS does not provide transactional isolation; open-trip queries are best-effort scans over the durable prefix and may exclude records appended after the scan advances beyond their indexed range.

We evaluate query performance using random time-range queries sampled from data spanning all three days, with query windows ranging from 1 s to the full dataset. Results are averaged over sampled queries for each modality and storage state. Both hot and cold queries are measured under cleared-cache conditions. Time to first byte (TTFB) measures the time from query start to the first returned record, including catalog lookup, trip selection, index scanning, and the first log access. Per-record latency measures time per iterator advance, including index progression, chunk boundary handling, header parsing, timestamp filtering, and cache/disk effects. Max RSS is the peak resident set size of the benchmarking
process, reflecting the transient footprint of mappings and buffers during retrieval.

\begin{table}[t]
\small
\centering
\caption{Hot query performance. For closed trips, results are measured using random time-range queries sampled from the full three-day SSD-resident dataset. For open trips, the query
is issued on the currently open trip while ingestion continues.}
\label{tab:hot-query-performance}
\resizebox{\columnwidth}{!}{%
\begin{tabular}{@{}c cc cc cc@{}}
\toprule
\multirow{2}{*}{\textbf{Metric}} 
& \multicolumn{2}{c}{\textbf{GPS}} 
& \multicolumn{2}{c}{\textbf{Image}} 
& \multicolumn{2}{c}{\textbf{Pointcloud}} \\
\cmidrule(lr){2-3}\cmidrule(lr){4-5}\cmidrule(lr){6-7}
& \textbf{Closed} & \textbf{Open} & \textbf{Closed} & \textbf{Open} & \textbf{Closed} & \textbf{Open} \\
\midrule
TTFB (ms) & 0.912 & 0.967 & 9.133 & 0.928 & 3.004 & 0.888 \\
\midrule
Per record latency avg (ms) & 0.002 & 0.001 & 0.811 & 0.006 & 0.696 & 0.006 \\
Per record latency p50 (ms) & 0.002 & 0.001 & 0.782 & 0.003 & 0.685 & 0.004 \\
Per record latency p95 (ms) & 0.005 & 0.003 & 0.831 & 0.011 & 0.772 & 0.012 \\
Per record latency p99 (ms) & 0.005 & 0.005 & 1.245 & 0.017 & 1.007 & 0.017 \\
\midrule
Max RSS (KB) & 46856 & 17152 & 349156 & 29352 & 453528 & 27656 \\
\bottomrule
\end{tabular}%
}
\vspace{-10pt}
\end{table}

Table~\ref{tab:hot-query-performance} shows that SSD queries return the first byte within 1--9 ms and keep per-record latency low. GPS is uniformly fast because many small records amortize per-chunk work. For larger modalities, open-trip queries are often faster because recently flushed chunks are likely page-cache resident; their smaller working set also lowers RSS.

\begin{table}[t]
\centering
\caption{Cold query performance over archived data spanning all three days. The reported results use random time-range queries drawn from archived trips across mixed day folders.}
\label{tab:cold-query-performance}
\begin{tabular}{@{}c c c c@{}}
\toprule
\textbf{Metric} & \textbf{GPS} & \textbf{Image} & \textbf{Pointcloud} \\
\midrule
TTFB (ms)                & 25.643  & 43.991   & 25.643   \\
\midrule
Per record latency avg (ms) & 0.003    & 4.889 & 1.718 \\
Per record latency p50 (ms) & 0.002    & 0.005    & 0.004    \\
Per record latency p95 (ms) & 0.006    & 23.981 & 8.403 \\
Per record latency p99 (ms) & 0.006    & 25.787 & 9.188 \\
\midrule
Max RSS (KB)             & 48836   & 34772   & 28048   \\
\bottomrule
\end{tabular}%
\vspace{-10pt}
\end{table}

Table~\ref{tab:cold-query-performance} shows that archived HDD queries remain directly accessible, with first-byte latencies below 44 ms without unpacking. Cold TTFB is higher because the first matching chunk must be fetched from the archive, and tail latency is dominated by chunk-boundary reads; peak RSS remains modest because retrieval uses transient chunk buffers.
\paragraph{Archive Behavior.}
We evaluate archival by migrating completed per-day logs from the SSD hot tier to the HDD cold tier. For each modality and day, append logs are packed into a single \texttt{.tar} archive prior to transfer. This converts many small objects into large sequential writes, reducing fragmentation and improving HDD throughput. Archival latency is defined as the end-to-end time from archive invocation to successful commit, including tar creation, HDD writeback, metadata updates, and SSD cleanup.

\begin{table}[H]
\centering
\caption{Archival performance over three days. Each run archives three modality folders.}
\label{tab:archive-performance}
\begin{tabular}{@{}c c c c@{}}
\toprule
\textbf{Metric} & \textbf{Day 1} & \textbf{Day 2} & \textbf{Day 3} \\
\midrule
HDD written (GB) & 24.59 & 26.66 & 25.39 \\
Wall time (s) & 227.12 & 247.29 & 237.20 \\
Throughput (MBps) & 108.27 & 107.81 & 107.04 \\
Avg RSS (MB) & 33.79 & 34.04 & 34.04 \\
\bottomrule
\end{tabular}
\vspace{-5pt}
\end{table}

Table~\ref{tab:archive-performance} summarizes archival over three days. Archived volume ranges from 24.59 to 26.66 GB, throughput stays near 107 MB/s, and average RSS stays near 34 MB. At this rate, one hour of driving data (about 24 GB) archives in roughly four minutes, giving a simple provisioning estimate for similar deployments.

\paragraph{Data Offload to Cloud.}
Data offload is evaluated with a TCP sender that retrieves one sensor topic and streams 100 or 1000 records over one persistent connection. Records are batched with \texttt{sendmsg}; we report throughput, payload goodput, and end-to-end latency averaged over 10 runs.

\begin{table}[ht]
\centering
\caption{Data offload TCP performance (reported number averaged over 10 runs and under wireless setting).}
\label{tab:data-offload-tcp}
\begin{tabular}{@{}c c c c c@{}}
\toprule
\textbf{\# Records} & 
\textbf{Data Type} & 
\begin{tabular}[c]{@{}c@{}}\textbf{Throughput}\\ \textbf{(MB/s)}\end{tabular} & 
\begin{tabular}[c]{@{}c@{}}\textbf{Goodput Ratio}\\ \textbf{(\%)}\end{tabular} & 
\begin{tabular}[c]{@{}c@{}}\textbf{Latency}\\ \textbf{(ms)}\end{tabular} \\
\midrule
\multirow{3}{*}{100}  & GPS        & 4.48 & 73.03  & 4.40 \\
     & Image      & 50.18 & 99.99 & 1020.87 \\
     & Pointcloud & 50.49 & 99.99 & 515.12 \\
\midrule
\multirow{3}{*}{1000} & GPS        & 5.58 & 74.77 & 15.31 \\
     & Image      & 51.02 & 99.99 & 10272.70 \\
     & Pointcloud & 51.92  & 99.99 & 4514.85 \\
\bottomrule
\end{tabular}%
\vspace{-5pt}
\end{table}

Table~\ref{tab:data-offload-tcp} shows near-perfect goodput for image and point cloud streams because batching amortizes protocol overhead. GPS has lower goodput because headers and checksums are a larger fraction of its small records. Latency scales mainly with payload volume.

\paragraph{Encryption Compatibility.}
Privacy-sensitive driving data may require encryption before persistence or export. AVS can apply authenticated encryption at chunk or archive granularity after reduction; in a 271-frame AES-256-GCM image append test, average latency was 7.52 ms, p95 was 9.45 ms, and maximum latency was 9.80 ms, within the 10 Hz camera period. A production deployment would still need full key-management and access-control support.
\section{Discussions and Limitations}
\label{paper:discussion}
The AVS prototype provides one concrete design point for queryable onboard retention, but the broader value of the system is in the tradeoffs it exposes. This section summarizes the main lessons from the design and evaluation, followed by limitations and open challenges for future vehicle data engineering.

\paragraph{Findings.}
First, retention should be treated as an ingest-time optimization problem. In conventional data systems, storage often acts as a passive backend, while expensive data processing is performed later as an offline task. However, autonomous vehicle streams contain strong spatial and temporal redundancy, and similar redundancy removal already appears in real-time driving pipelines. For long-horizon onboard storage, applying lightweight, modality-aware reduction before persistence can reduce storage pressure, lower downstream processing cost, and make retained data more usable for later applications. This shifts storage from a passive recorder to an active data management layer that decides what information is worth preserving. 

Second, hierarchical vehicle storage must preserve queryability across tiers. Tiered storage is a natural response to limited onboard SSD capacity, but directly applying conventional hot/cold archival is insufficient for vehicle histories. Once data is moved to HDD, naive archival can turn retained sensor streams back into opaque files that require extraction or replay before use. But the archived data may still need to be quickly and selectively accessed by applications on board. This reframes the hierarchy from a backup mechanism into a continuous queryable retention substrate across hot and cold vehicle data.

Third, vehicle storage may need stream-oriented durability semantics rather than full database-style transactions. AV sensor data is continuous, high-rate, and append-dominant, and power loss can occur before all buffered records reach persistent storage. In this setting, the key durability goal is not only to commit every individual record transactionally, but also to recover quickly to a consistent history with a clear corruption boundary.

\paragraph{Limitations and Open Challenges.}
The first limitation is the generality of the current reduction policy. AVS selects fixed reduction parameters mainly through dataset-based experiments. This validates one feasible design point, but broader datasets, scenes, and downstream tasks are needed to understand transferability. The deeper challenge is not just choosing one better threshold. Different applications such as safety analysis, model debugging, and forensic inspection may not tolerate the same information loss. Future work should study utility-aware reduction policies that jointly optimize reduction ratio, downstream usability, and real-time feasibility.

The second limitation is prototype scale. AVS evaluates concurrent ingest from one LiDAR, one camera, and one GPS stream. Although the prototype shows bounded memory use at this scale, production vehicles may include more cameras, higher-resolution LiDAR, radar, ultrasonic sensors, inertial sensors, and richer vehicle bus data. As sensor count and rate increase, memory pressure, CPU cost, metadata size, and storage wear will also increase. This creates an open challenge for both data engineering and storage communities: how can we design bounded-memory ingest, indexing, and reduction mechanisms for continuously growing multimodal streams? At the storage layer, future work should also investigate how to improve device lifetime under sustained vehicle writes.

Finally, AVS does not yet address privacy, security, and governance. Vehicle histories may contain faces, license plates, home locations, driving routines, and behavior traces. As vehicles become more connected and computationally powerful, they will increasingly resemble mobile data platforms, but with stronger safety and physical-world implications. Future systems need vehicle-specific threat models, access control, anonymization, auditing, and policy-aware query mechanisms. This is an area where the community can contribute substantially by connecting stream processing, edge data management, privacy-preserving analytics, and domain-specific governance.
\section{Conclusion}
\label{paper:conclusion}
Autonomous vehicle data are large, heterogeneous, and increasingly useful beyond real-time driving pipelines. AVS identifies an emerging onboard data-management operating point: transforming transient sensor streams into retained, queryable histories under embedded deployment constraints. This turns vehicle data management into a full-lifecycle problem, spanning ingest-time reduction, durable logging, tiered retention, recovery, and selective query access.

This paper presents AVS, an onboard vehicle data-management system with a vehicle-specialized append-only logger as its core storage mechanism. AVS combines utility-aware preprocessing, chunk-indexed logging, and query-preserving hot/cold archival to support sustained ingest, bounded recovery, and selective access on real vehicle hardware. The prototype demonstrates one feasible design point for onboard vehicle data engineering, while exposing open questions in adaptive retention, larger-scale multimodal ingest, storage endurance, cold-tier query processing, and privacy-preserving governance. AVS is only an initial step, and we hope this work motivates further research across data engineering, storage systems, and autonomous driving toward the next generation of vehicle computing data systems.






\bibliographystyle{IEEEtran}
\bibliography{reference}

\end{document}